\documentclass[a4paper,twocolumn,aps,prd,nolongbibliography,superscriptaddress,showpacs,showkeys,amsmath,amssymb,floatfix,nofootinbib]{revtex4-1}

\usepackage{graphicx}
\usepackage{amssymb}
\usepackage{lmodern}

\usepackage[T1]{fontenc}
\usepackage[utf8]{inputenc}
\usepackage[dvipsnames]{xcolor}
\usepackage[unicode=true,pdfusetitle,
 bookmarks=true,bookmarksnumbered=true,bookmarksopen=true,bookmarksopenlevel=1,
 breaklinks=false,pdfborder={0 0 0},backref=false,colorlinks=true]
 {hyperref}
\hypersetup{
 citecolor=blue,filecolor=blue,linkcolor=blue,urlcolor=blue}
\makeatletter

\usepackage{textcomp}
\usepackage[caption=false]{subfig}
\usepackage{slashed}
\usepackage{braket}
\usepackage{braket}
\usepackage{graphicx}
\usepackage{float}
\pdfpageheight\paperheight
\pdfpagewidth\paperwidth

\renewcommand{\[}{\begin{equation}}
\renewcommand{\]}{\end{equation}} 

\usepackage{array}
\usepackage[normalem]{ulem}

\makeatother

\begin{document}

\title{White dwarfs and revelations}

\author{Ippocratis D. Saltas} 
\email{ippocratis.saltas@fzu.cz}
\affiliation{CEICO, Institute of Physics of the Czech Academy of Sciences, Na Slovance 2, 182 21 Praha 8, Czechia}

\author{Ignacy Sawicki} 
\email{ignacy.sawicki@fzu.cz}
\affiliation{CEICO, Institute of Physics of the Czech Academy of Sciences, Na Slovance 2, 182 21 Praha 8, Czechia}

\author{Ilidio Lopes} 
\email{ilidio.lopes@tecnico.ulisboa.pt}
\affiliation{Multidisciplinary Center for Astrophysics \& Department of Physics,\\ Instituto Superior T\'ecnico, University of Lisbon, Av. Rovisco Pais 1, 1049-001 Lisboa, Portugal}

\begin{abstract}
We use the most recent, complete and independent measurements of masses and radii of white dwarfs in binaries to bound the class of non-trivial modified gravity theories, viable after GW170817/GRB170817, using its effect on the mass-radius relation of the stars. 
We show that the uncertainty in the latest  data is sufficiently small that residual evolutionary effects, most notably the effect of core composition, finite temperature and envelope structure, must now accounted for if correct conclusions about the nature of gravity are to be made. 
We model corrections resulting from finite temperature and envelopes to a base Hamada-Salpeter cold equation of state and derive consistent bounds on the possible modifications of gravity in the stars' interiors, finding that $Y< 0.14$ at 95\% confidence, an improvement of a factor of three with respect to previous bounds. Finally, our analysis reveals some fundamental degeneracies between the theory of gravity and the precise chemical makeup of white dwarfs.
\end{abstract}
\maketitle

\section{Introduction}
The paradigm of General Relativity (GR) has been brought under intense scrutiny in the last few years, mainly due to the discovery of the accelerating Universe and the lack of a theoretical explanation of the value of a cosmological constant implied by it \cite{Weinberg:1988cp}. This sparked off the investigation of dynamical mechanisms to explain the observed acceleration at large scales, frequently involving theories of gravity beyond GR (For a review see \cite{Clifton:2011jh}). These dynamical models introduce new dynamical degrees of freedom in the gravitational sector, the simplest case being that of a scalar field interacting minimally with the spacetime metric. In the case where the new field interacts non-trivially with curvature, it inevitably leads to a modification of gravity, which leaves an imprint at local scales through a fifth-force effect. The most popular theories in this regard have been various flavours of scalar-tensor models. They are subsumed in the so-called Horndeski scalar-tensor theories \cite{Horndeski:1974wa} and their more recent extension known as {\it Beyond Horndeski} models \cite{Zumalacarregui:2013pma,horndeski} and \emph{Degenerate Higher-Order Scalar-Tensor} (DHOST) theories \cite{BenAchour:2016fzp,Langlois:2017mxy}, which correspond to the most general theories that can be constructed from the metric and a scalar field which are free from Ostrogradski instabilities. These theories generalise the archetypal models of Brans-Dicke/$f(R)$ gravity with the corresponding scalar field-curvature interactions operating in a {\it non-trivial} manner. They can quite easily model the acceleration of the expansion of the universe, without involving a cosmological constant and have quite a rich phenomenology in structure formation, the constraining of which is the target of the current and upcoming cosmological surveys. 

It is important to point out that modified-gravity theories, in addition to modifying the forces between massive bodies, also can modify the propagation of gravitational waves (GW), including their speed. The recent measurement of the GW speed by the LIGO/VIRGO collaboration together with Fermi/Integral (GW170817 with its electromagnetic counterpart GRB 170817 \cite{TheLIGOScientific:2017qsa, Monitor:2017mdv}) to be essentially equal to that of light, has heavily constrained the allowed theory space of modifications of gravity. In the case of scalar-tensor theories, the scalar is now allowed to be at most conformally coupled to curvature \cite{Ezquiaga:2017ekz,Creminelli:2017sry,Sakstein:2017xjx,Baker:2017hug}.

A crucial feature of theories beyond GR is the existence of some screening mechanism with the role of suppressing potential fifth-force effects mediated by the scalar field inside or sufficiently close to matter sources such as stars and galaxies, allowing them to evade in this way tests of gravity in the Solar System. However, the remaining, non-trivial%
\footnote{The term ``non-trivial'' here denotes those theories where the scalar field couples to curvature beyond the standard Brans-Dicke type of coupling, as for these cases, the Ostrogradski instability is avoided in a highly non-trivial manner.} %
and viable scalar-tensor theories after GW170817 and GRB 170817 present a challenging prediction: Through a partial breaking of the screening mechanism within massive sources, they predict a departure from Newtonian gravity \emph{inside} the source, allowing for a scale-dependent fifth-force, with standard gravity only to be recovered outside the object \cite{Kobayashi:2014ida}. This would affect the equilibrium structure of the star and in turn imply a direct phenomenological impact on scaling relations such as the mass-radius relations, which is what will concern us here. Note that, in addition to the above, after GW170817, the modification of gravity we consider here is the only one that can weaken gravity in cosmology without pathological instabilities \cite{Amendola:2017orw}.

In this work, we will exploit this departure from the standard paradigm inside massive sources to understand the effect of modifications of gravity on white dwarfs (WD). The choice of WDs as stellar laboratories for gravity has been a popular one in the literature, with the reason usually relating to simplicity: Although compact enough, white dwarfs lie within the validity of the Newtonian approximation, while their equation of state is relatively well-understood, in comparison with more compact objects such as neutron stars.%
\footnote{For a review on the physics of WDs we refer the reader to Refs~\cite{WDphysics, Shapiro}.} %

\noindent{\it Let us summarise the goals of the work as follows:}

We predict the mass-radius relation for WDs within the remaining, viable theory space of scalar-tensor theories after GW170817, starting from the refined version of the standard Chandrasekhar model, in the presence of its leading-order (Coulomb) corrections at zero temperature according to Hamada-Salpeter (HS). We show how the cold HS description fails in view of the high precision of latest data set, and we proceed advancing our modelling of the WD through the implementation of {\it finite temperature} and {\it envelope} corrections, employing results from publicly available stellar-structure simulations in GR. This allows us to sufficiently reduce systematic effects and produce constraints robust enough to be able to constrain modifications of gravity, improving limits from WD physics by a factor of 3. As an aside, our analysis provides with a first understanding of the subtle degeneracies between the fine structure of the star's interior and the possible gravity modification.


\section{Gravitational equations in the star's interior  \label{sec:Hydro}}

Let us start by introducing the necessary theoretical framework for our subsequent analysis. Scalar-tensor theories, are full relativistic theories, just like GR, based on a space-time metric.  In WD stars, the mass is essentially all provided by atomic nuclei, which remain non-relativistic. One can therefore describe the gravitational law in the star's interior using a (modified) Poisson equation relating the Newtonian potential $\Phi$ with the mass density $\rho$, neglecting the other gravitational potential and the pressure from relativistic electrons. In this limit, and for the remaining, viable and non-trivial modified gravity theories after GW170817 the screening mechanism is violated inside massive bodies, and the Newtonian potential $\Phi$ is sourced through an extra term in addition to the standard GR one, describing the modification of gravity \cite{Kobayashi:2014ida,Crisostomi:2017lbg,Dima:2017pwp} 

\begin{align}
\nabla^2 \Phi = 4 \pi G \rho + G \frac{Y}{4} \nabla^2 \left( \frac{dM}{dr} \right), \label{Poisson}
\end{align}
with $G$ Newton's constant, $M=M(r)$ the mass contained in a sphere of radius $r$ from the centre of the star, and $Y$ a dimensionless coupling constant controlling the strength of the modification. The derivative $dM/dr$ is expressed in terms of the density through the usual mass-conservation equation,
\begin{align}
\frac{dM(r)}{dr} = 4 \pi r^2 \rho(r). \label{Hydro2}
\end{align}
Notice that, since $d\rho/dr < 0$ within the star's interior, $Y > 0$ will act as to weaken gravity, while the opposite is true for $Y < 0$. In principle the value of $Y$ is not fixed in the universe, but can be dependent on cosmological time and the environment of the star. However, it should be approximately constant inside the Milky Way, where all the WDs in the dataset we will use are located.

We will be assuming that the star is in hydrostatic equilibrium, i.e.\ that gravity perfectly balances the interior pressure of the star, which translates to $\nabla \Phi = - \nabla P$, and under the assumption of spherical symmetry, equation (\ref{Poisson}) yields
\begin{align}
\frac{dP}{dr} = - \frac{GM(r)}{r^2}\rho(r) -  \frac{G \cdot Y}{4}\frac{ d^{2} M(r)}{d r^2} \rho(r). \label{Hydro1}
\end{align}
Equations (\ref{Hydro1}) and (\ref{Hydro2}) form a closed system when provided an equation of state relating $P = P(\rho)$ and with the initial conditions at the center of the star as $P = P_c$ ($\rho = \rho_c$), $M = 0$ and $P'(r =0) = M'(r=0)$ with $' \equiv d/dr$. It is interesting to notice that equation (\ref{Hydro1}) can be re-cast into its standard form, but with an effective scale-dependent Newton's constant, 
\begin{equation}
\frac{G_{\text{eff}}}{G} = 1 + \frac{Y}{4}\frac{r^2}{M(r)} \frac{d^{2} M(r)}{dr^2}. \label{Geff}
\end{equation}

Considering that the mass contained within a sufficiently small shell around the center of the star can be approximated as $M(r) \approx (4 \pi/3)\rho_c r^3$, from (\ref{Geff}) it turns out that sufficiently close to the center it is $G_{\text{eff}}/G \approx 1 + (3/2) Y$. Stability reasons require that $G_{\text{eff}} > 0$, which in turn implies that $Y \geq -2/3$ \cite{Saito}. In fact taking into account relativistic effects, this bound can be shown to be stronger, with $Y>-4/9$ \cite{Babichev:2016jom}. This places a lower, theoretical prior on $Y$, since for lower values spherically symmetric solutions cease to exist.

Let us now assume that we are given an equation of state relating $P =P(\rho)$, with $P = P(x)$ and $\rho = \rho(x)$ parametrically defined through the dimensionless Fermi momentum $x \equiv p_F/(m_\text{e} c)$ (see next section). We can then express the hydrostatic equation (\ref{Hydro1}) as a differential equation for $x = x(r)$ as
\begin{align}
\frac{d x}{dr} = - G \rho\left(  \frac{M}{r^2} + 2\pi Y \cdot \rho r\right)\cdot \left( \frac{dP}{dx} + \pi Y G \cdot \rho r^2  \frac{d\rho}{dx} \right)^{-1}. \label{Hydro3}
\end{align}
Equations (\ref{Hydro3}) and (\ref{Hydro2}), together with an appropriate equation of state are the system of hydrostatic equations we solve. Note that, for a typical central pressure in a WD, one can estimate from $P_c \sim G M^2/R^4$ that $x_{c} \sim \mathcal{O}(1)$. 

The numerical procedure to solve the hydrostatic equations is as follows: Setting our initial conditions at the center with $r_c = 10^{-30} R_{\odot}$, $M_c =10^{-30} M_{\odot}$, we construct a family of solutions using an appropriate implicit-integration solver, iterating over $x_c \in [0.5, 7]$ (at the centre) with a step size $\delta x = 0.01$, and $Y \in [-0.66, 3]$ with $\delta Y = 0.036$ respectively. Since in principle the equation of state will depend on core composition through the relevant atomic number ($Z$), we iterate above procedure for $Z = 4, 6, 8$ corresponding to helium, carbon and oxygen cores respectively. This gives a density profile $\rho = \rho(r)$, and we choose the radius $R$ of the star as the root where $x(r = R) = 0$ within some precision, while  its mass as $M = M(r = R)$. Given a choice of equation of state, this provides us with a grid of solutions for the WD's radius and mass for each choice of central density, composition and $Y$. We can then proceed with comparing these predicted stars with the available observational data, as we discuss in Section~\ref{sec:data}.

\section{The Chandrasekhar model and beyond \label{sec:eos}}

The choice of the equation of state, and the modelling of the different layers that make up the star's interior is fundamental in any analysis of stellar structure. As we will show, uncertainties in its finer details can significantly interfere with constraints on the underlying law of gravity. In this section we discuss the standard paradigm for the equation of state for WDs and its refinements. 

The earliest equation of state for WDs is Chandrasekhar's model \cite{Chandra1}, which assumes that the gravitational force is balanced solely by the degeneracy pressure of the spherically symmetric electron gas in the star's interior (see \cite{WDphysics}). This has been in fact the most popular choice in the study of modified gravity models with white dwarfs in the literature (see e.g.\ \cite{Koyama, Bertolami:2016ylu, Jain:2015edg}). It can be shown that electron pressure for a non-interacting electron gas is given by \cite{Chandra1, WDphysics, Shapiro}
\begin{align}
P_{0} = \frac{1}{24 \pi^2} \cdot \frac{m_{e}^4 c^{5}}{\hbar^3} \cdot \left[ 3 \text{sinh}^{-1} x  + x(2x^2 -3)(x^2 + 1)^{1/2} \right], \label{P_0}
\end{align}
with the dimensionless Fermi momentum $x \equiv p_{F}/(m_{e} c)$. The limits $x \ll 1$ ($x \gg 1$) yield the well-known (non-) relativistic limits for the equation of state, reducing it to the simpler polytropic form, with $P \sim \rho^{(n+1)/n}$ with $n$ the polytropic index. The electrons can be safely assumed to provide a negligible contribution to the energy density.  We take the mass to be that of the ionised nuclei the distribution in space of which follows the electrons,
$
\rho \simeq (9.7395\cdot 10^5)\frac{A}{Z}  x^3 \text{g}/\text{cm}^3,
$
where $A$ is the atomic number of the core's element. It is important to notice that, there is no explicit dependence on $Z$, with $A/Z = 2$ for C, O, and He, hence the model is degenerate with respect to the core composition.

Although the Chandrasekhar model captures the fundamental features of the star's equation of state, it is nevertheless a zeroth-order approximation. A first step towards a more realistic description, is giving up on the uniform distribution of ions and electrons and assuming their distribution as a perfect lattice, according to Hamada-Salpeter \cite{Salpeter, HS}. In this setup, the leading-order correction to the pressure of the free-fermion model is the Coulomb interactions between the electron gas and the positive ions. The emergent non-uniform distribution of the electrons is taken into account by the next-to leading order correction known as  Thomas-Fermi, while further sub-leading corrections are due to electron-electron interactions (exchange energy and correlation correction respectively). The relevant improvements for above corrections to the pressure are provided in \cite{Salpeter} and the model is known as the zero-temperature Hamada-Salpeter model (HS). Here we shall only provide the scaling of the corrections to the pressure with $x \equiv p_F/(m_e c)$ and atomic number $Z$, ignoring (dimensionful) pre-factors. They read as \footnote{The expression for the exchange energy is rather lengthy and we refer to \cite{Salpeter} for more details. Further sub-leading corrections described in \cite{Salpeter} we omit here.}
\begin{align}
P_{\text{C}},\; P_{\text{TF}}, \; P_{\text{Correl.}} \sim Z^{2/3} x^4, \;  \frac{ Z^{4/3}x^5}{(1+x^2)^{1/2}}, \; x^3.
\end{align}
The above corrections are understood to be additional to the zeroth-order term $P_0$ (\ref{P_0}), with the matter energy density still described by its non-relativistic expression provided earlier. They result in a new attractive interaction, reducing the radius of a white dwarf at a fixed mass by $4-5\%$ in the relevant mass range in general relativity.

\begin{figure}[t]
	\includegraphics[width=\columnwidth]{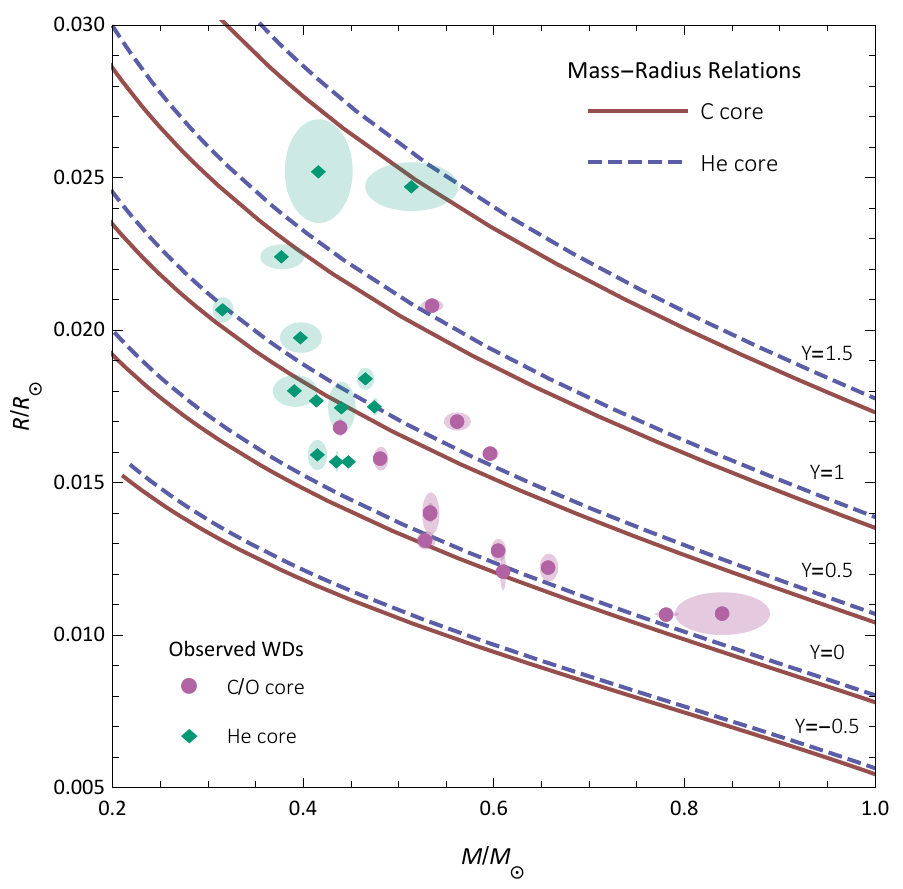} 
	\caption{Theoretical mass-radius curves based on the zero temperature Hamada-Salpeter equation of state, for different values of the modified-gravity parameter $Y$ and core composition. The data points correspond to the observed data, with the ellipses showing the $1\sigma$ observational uncertainties, while their core-classification is extracted with $Y = 0$ according to \cite{Parsons2017}. It can be seen that $Y < 0$ acts as to make stars smaller, and the opposite for $Y > 0$. The stars departing the most from the zero-temperature curve, towards the upper left part of the plot, correspond to those with higher effective temperatures. This is an important aspect that needs to be accounted for when constraining models beyond GR against observations.
		\label{plot:MR}}
\end{figure}

In this context, we have implemented the HS equation of state for the pressure in deriving the theoretical prediction for the star's mass/radius profiles, following the numerical procedure outlined in the previous section (see the discussion after Eq.~(\ref{Hydro3})). We believe that this is explored for the first time in the literature within the context of theories beyond GR. We present the result of our calculations --- mass-radius relations for white dwarfs in modified gravity --- in Fig.~\ref{plot:MR}. As indicated earlier, a positive $Y$ weakens gravity and indeed we find that, for a fixed mass, the expected radius of the zero-temperature white dwarf is larger.

\section{Temperature/envelope matters} \label{sec:Temp/Env}

Treating the WDs as above, with the HS equation of state at zero temperature, leaves out important physics that turns out to be the dominant systematic effect. 

The data relevant for our purposes reduce to the mass, radius and temperature of each WD. If we neglect the temperature, for stars in the realistic mass-radius range, it is always possible to find some {\it local} value of $Y$ which will give the correct mass-radius relation at the individual star's mass for essentially any equation of state. On the other hand, a {\it global} $Y$ parameter can be constrained by confronting the full data set with the overall scaling (shape) of the mass-radius relation, provided the systematic effect of the neglected physics is uncorrelated and sufficiently small compared to the precision of the observations. We show in Section~\ref{sec:data} that the the data set we use is now sufficiently precise that this is no longer the case and leads to spurious constraints on $Y$. 

\begin{figure*}[htb]
	\includegraphics[width=2\columnwidth]{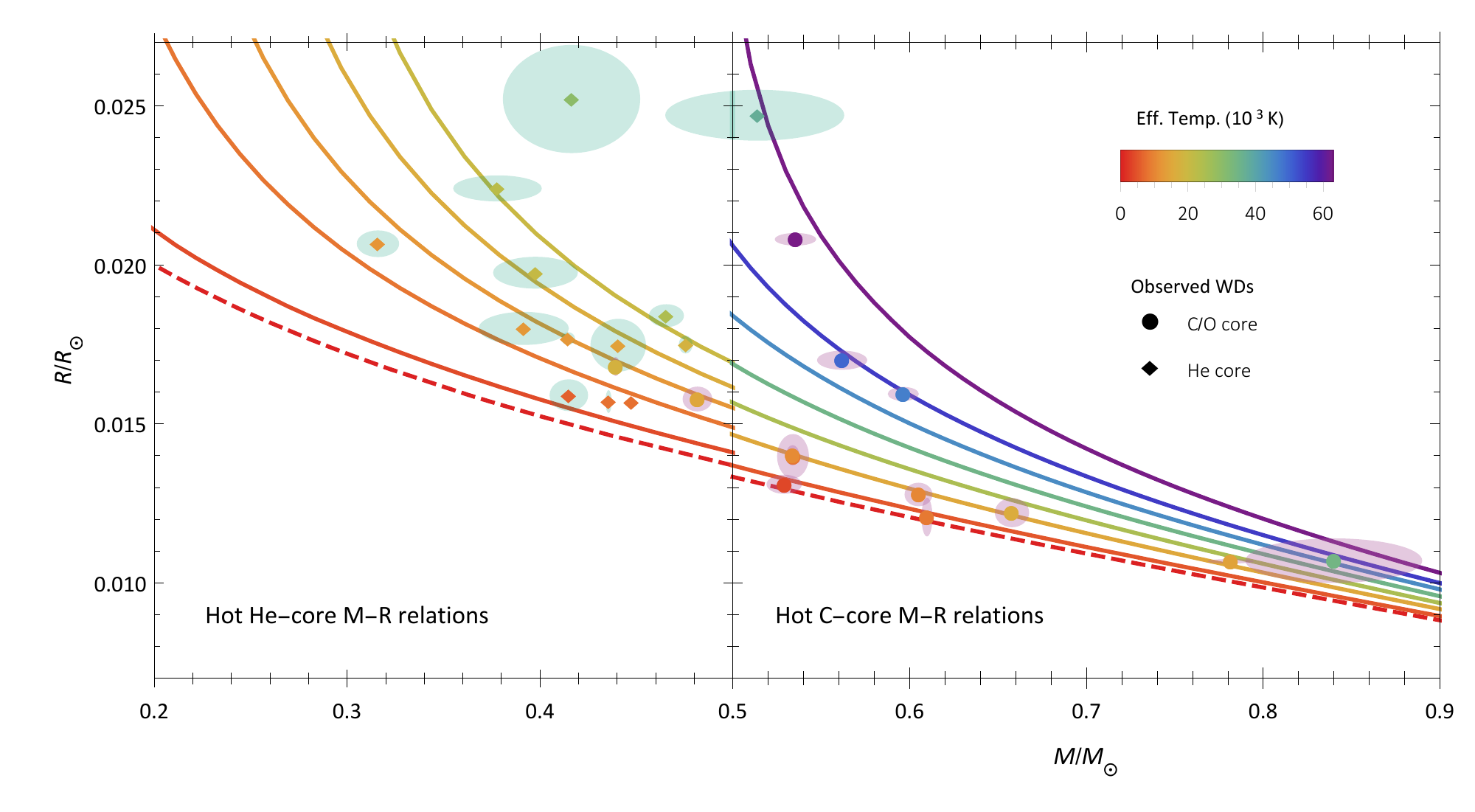} 
	\caption{Mass vs radius for the white dwarf sample of \cite{Parsons2017} with 1$\sigma$ errors. Core composition as identified \emph{op.\ cit.} Dashed lines are the mass-radius curves for the HS equation of state in GR. Solid lines correspond to mass-radius curves for various effective temperatures as predicted by full simulations of stellar evolution in GR according to \cite{Panei:1999ji}. In the left panel, the curves are for He-core white dwarfs with an H envelope with mass fraction $3\cdot 10^{-4}$. In the right panel, the curves represent mass-radius relations of C-core white dwarfs, with a He envelope with mass fraction $10^{-2}$ and a H envelope with mass fraction $10^{-5}$. Color coding represents effective temperatures. \\
		For the most massive C-core WDs, the temperature effects on the radius are less than 10\%. However, this rapidly increases for lighter WDs, especially for He cores, to reach 40\% for the range of temperatures present in the catalog of \cite{Parsons2017}. Neglecting the temperature/envelope effects introduces a significant systematic error and leads to a false interpretation that the stars require gravity to be modified. \label{plot:MRT}}
\end{figure*}

A realistic WD still possesses an internal temperature, meaning that the equation of state is not barotropic, i.e. we have $P =P(\rho,T)$, and its modelling requires precise simulations of the star's evolution from its progenitor to the WD stage.  In Fig.~\ref{plot:MRT}, we present the data of the catalog of \cite{Parsons2017}, along with the publicly available families of mass-radius relations parametrized by the effective temperature, generated with such a simulation for Ref.~\cite{Panei:1999ji}.%
\footnote{For similar studies see for example \cite{LambHorn1975, Althaus1997,Althaus1998,Panei2007}.} %
The fact that the WDs in Fig.~\ref{plot:MRT} lie close to the lines corresponding to their effective temperature implies that the deviation of the theory of gravity from GR should be small. To properly quantify this, we need to take the temperature effect of the mass-radius relation into account. As the star evolves along its cooling track, the main gravo-thermal process is the release of thermal energy accompanied by a phase of contraction.  While for the most massive stars the effect of the cooling is small, less than a 10\% decrease in the radius for the sort of temperatures in the sample, it can be as large as 40\% for the lightest stars, especially for those with helium cores.%
 \footnote{Our discussion is in qualitative accordance with early studies of temperature effects on the star's radius in the context of GR, see e.g.\ \cite{Hubbard1970}.} %
In all the cases, the radii of the hot stars are larger than those of the zero-temperature objects, a property we will exploit to put a bound on the modification of gravity.

Yet another important residual evolutionary feature of WDs is the existence of an envelope of non-degenerate matter surrounding the star's degenerate core. Despite the fact that it usually comprises less than 1\% of the star's mass, its effect on the actual radius of the star can be significant, with thicker envelopes yielding larger stars at the same mass and temperature. We use ``thickness'' to refer to the envelope's mass fraction of the total WD mass, rather than necessarily the physical extent.

The thickness and chemical composition of the envelopes depends on the interplay of the delicate processes during the red giant branch phase, and element diffusion as a function of time after the star has evolved into a WD. An accurate description for the envelope composition and thickness requires therefore adequate modelling, with a careful accounting for the diffusion process. 

Precision simulations of this sort, including element diffusion for C/O-core WDs with envelopes have been performed in Ref.~\cite{Althaus2010} and further discussed in Ref.~\cite{Romero2012} within an asteroseismology context, while for He-core WDs, such simulations were presented in Ref.~\cite{Althaus1998}, where a wide range of possible thickness of hydrogen envelopes are studied. As it turns out, the fraction of helium in the envelope is rather insensitive to the prior evolutionary details, while that of hydrogen can vary significantly. 

From an observational viewpoint, inferring the envelope's mass fraction presents a challenging issue if it is possible at all. While measurements of the star's mass/radius are not very informative in this regard, asteroseismological observations can provide a powerful tool for this purpose. In this context, the observed pulsations of various C/O-core WDs, known as ZZ Ceti stars, in a particular temperature range and characteristic timescales, have allowed for partial confirmation of the simulated model parameters and predictions \cite{Romero2012}. The observation of the possible pulsations for the stars of our data set would certainly provide an important complementary observable able to break parameter degeneracies when constraining modifications of gravity.

The important point in this regard is that simulations point to a minimum and maximum thickness for both helium and hydrogen envelope for C/O-core WDs and therefore in principle would allow for an estimate of the range of the possible systematic effect from the unknown envelopes.\\

No simulations of WD evolution are publicly available within modified gravity. We instead assume that the effect of temperature and envelope on the radius is additive and does not significantly change for the kind of values of $Y$ that are consistent with the data. We use the publicly available results of the WD evolution simulations performed in Ref.~\cite{Panei:1999ji} to extract the corrections to the radius, $\delta R$, for each mass, temperature, core and envelope type, and we interpolate between the available simulated data to cover the mass/temperature values suggested by our dataset. The relevant corrections are then added on top of the cold HS model computed in a modified gravity context. 

The validity of the above approximation depends on the size of $Y$, but is a correct approximation for sufficiently small deviations from GR. 
The effect of the envelope should be independent of the theory of gravity since the interaction between the two can only depend on $GM/R$ and, sufficiently close to the star's surface, $G_{\text{eff}} \simeq G$ (see equation (\ref{Geff})) is a constant. On the other hand, the equation of state depends on the temperature in the core and this has a highly non-trivial interaction with the core profile through the hydrostatic equation~\eqref{Hydro1} that can only be modelled through numerical simulation. But, for sufficiently low temperatures, the modified HS profile will be a good approximation. Essentially, we have assumed for the correction to the radius that $R \simeq R_{\text{HS}}(Y) + \delta R(M, T_{\text{eff}};Y=0)$, for the kind of $Y$ the data allow.

%
%

\section{The data set and statistical analysis \label{sec:data}}
For our statistical analysis we use a recently published catalog of 26 white-dwarf masses and radii along with their effective temperatures, presented in \cite{Parsons2017}, ranging between $M \sim 0.3 - 0.8 M_{\odot}$, $R \sim 0.01 - 0.025 R_{\odot}$ and $T \sim 7500 - 63000$~K respectively.%
\footnote{The catalogue is a compilation of 16 new measurements along with 10 more previous ones, taken from \cite{Bours2014, Parsons2012, Parsons2010,Parsons2016,Parsons2012b,Parsons2012c,Pyrzas2012,OBrien2001}. We notice also that, the previous similar catalogue of \cite{holberg} partly relies on an assumption on the mass-radius relation for the modelling of the stars' atmospheres.} %
These WDs are all members of eclipsing binaries, which allows for the \emph{independent determination} of mass, radius and effective temperature, without reference to a model for the WD interior. This makes this catalog unique: typically the mass-radius relation is assumed in the construction of the WD catalog and only one of these is truly independent. This allows us to use the catalog of Ref.~\cite{Parsons2017} to constrain modifications of gravity without introducing such biases. In addition, the average errors on the mass and radius in this sample are $3\%$, a significant improvement compared to, for example, Ref.~\cite{holberg} with average errors of $13\%$. It is this reduction in uncertainty that drives the unearthing of systematic effects related to temperature, as we discuss below.

Within GR, given a measurement of the mass, radius and effective temperature, the core composition and the thickness/composition of the envelope is inferred in a statistical manner with the aid of fully-featured simulated mass-radius curves. In particular, the core structure cannot always be inferred unambiguously, except for special cases such as very low-mass WDs in binaries, the evolutionary track of which dictates a core made of helium. Therefore, in the absence of some guiding physical argument, above stellar parameters can therefore be treated as nuisance parameters in any study of fundamental physics. 

For our analysis, we construct a likelihood for each star $i$, assuming no covariance between the reported errors on the observed mass $M^i_\text{obs}$ and radius $R^i_\text{obs}$, $\mathcal{L}_i \propto \text{exp}(-\chi_{i}^2/2)$ with
\begin{align}
\chi^2_{i} = \frac{(R_{\text{obs}}^{i}-R_{\text{th}})^2}{\sigma_{i R}^2} + \frac{(M_{\text{obs}}^{i}-M_{\text{th}})^2}{\sigma_{i M}^2}, \label{Chi2}
\end{align}
where $M_\text{th}$  and $R_\text{th}$ are the mass and radius derived assuming one of the equations of state we will discuss below, and the modified gravity parameter $Y$. The likelihood then depends on unobservable nuisance parameters such as the central Fermi momentum $x_c$ or the composition. We pick an appropriate prior for these and marginalize to obtain a posterior for the parameter $Y$ as given by each star. We discuss how we obtain constraints on the global $Y$ in Section~\ref{sec:Yconstr}.\\

\paragraph*{Hamada-Salpeter:} The HS equation of state provides us with full analytic control in modelling the interior structure of the WD in modified gravity. (See Section \ref{sec:Hydro} for a detailed description.) The solutions for the mass/radius profiles are parametrised by the unobservable central Fermi momentum $x_c$ and the atomic number of the core, $Z$. The effect of $Z$ is small (a few percent, see Fig.~\ref{plot:MR}), so we marginalize the posterior over He, C and O cores with an equal probability prior, which also reflects the reality that all of the species are present in some fraction in a heavier-cored star. This gives a two-dimensional likelihood on the $Y$-$x_c$ plane for each of the 26 stars. We find that these likelihoods are disjoint and this is not improved by marginalizing over the unobservable $x_c$ with a flat prior over the range $0.5<x_c<7.0$.

\begin{figure}[tb]
	\centering
	\includegraphics[width=\columnwidth]{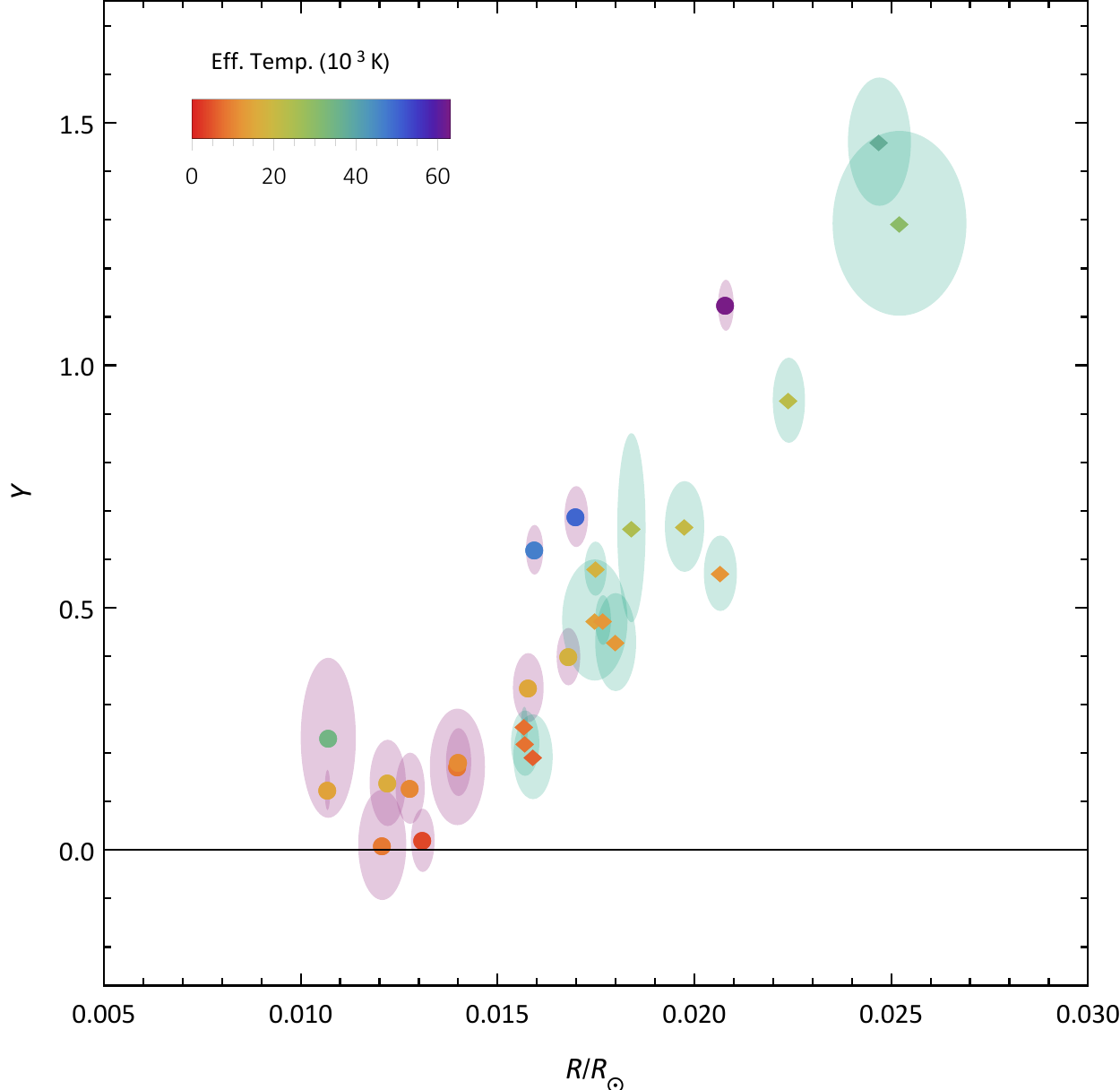}
	\caption{Preferred central values of the modified-gravity parameter $Y$ along with their 1$\sigma$ error against observed radii for each of the 26 stars of the dataset \cite{Parsons2017}, assuming the zero-temperature HS equation of state and having marginalised over the central density parameters $x_0$ and compositions $Z$ for each star. The data are in tension at least at 2$\sigma$ with both GR ($Y=0$) and each other: the stars' preferred $Y$ value clusters in non-overlapping groups. What is more, a clear (misleading) correlation between $Y$ and the star's radius emerges.\\
	We have colour-coded the stars according to their observed effective temperature: there is a clear trend between the temperature and the preferred value of $Y$, albeit with two independent series grouped according to the core compositions classification as per \cite{Parsons2017}. We thus see that temperature emerges as a systematic that must be taken into account in the modelling of the stars.\label{plot:YRT}} 
\end{figure}

Figure \ref{plot:YRT} shows the preferred central values and 1$\sigma$ errors for this marginalised posterior for $Y$ for each star in the data set. Most stars are not only incompatible with GR ($Y=0$) at the 2$\sigma$ level, but they are even incompatible with each other. When $Y$ is plotted against the stars' radius, there in fact appears a linear relationship between parameters, with larger stars seemingly preferring weaker gravity (larger $Y$). The tension means that one should not use this sample to put constraints on the common value of $Y$ in the Milky Way, but rather must improve the modelling. This bug is inherent to constraining gravitational theories with zero-temperature models of WDs, which have been widely used in the literature in this context. 

\paragraph*{Finite Temperature:} Here we include corrections from the finite temperature and envelope as discussed in Section~\ref{sec:Temp/Env}. To model temperature corrections to the HS radii, we use six series of results from simulations from Ref.~\cite{Panei:1999ji}: helium core ($0.18 \lesssim M/M_\odot \lesssim 0.50$ with $4\cdot 10^3$~K$<T<20\cdot 10^3$~K), carbon and oxygen cores ($0.30\lesssim M/M_\odot \lesssim 1.18$ and $5\cdot 10^3$~K$<T<145\cdot 10^3$~K) with two types of envelopes each: ``thin'' and ``thick''. The former case corresponds to He core stars possessing no envelope at all, and C/O ones a helium envelope ($M_{\text{He}}/M = 10^{-2}$), while the latter  to the case where He-core stars have a hydrogen envelope with $M_{\text{He}}/M = 3 \cdot 10^{-4}$, and C/O ones an envelope composed of helium and hydrogen ($M_{\text{H}}/M  =10^{-5}$ and $M_{\text{He}}/M =10^{-2}$ respectively). 
\begin{figure*}[htb] 
	\subfloat[Thin envelopes\label{plot:YRthin}]
	{\includegraphics[width=\columnwidth]{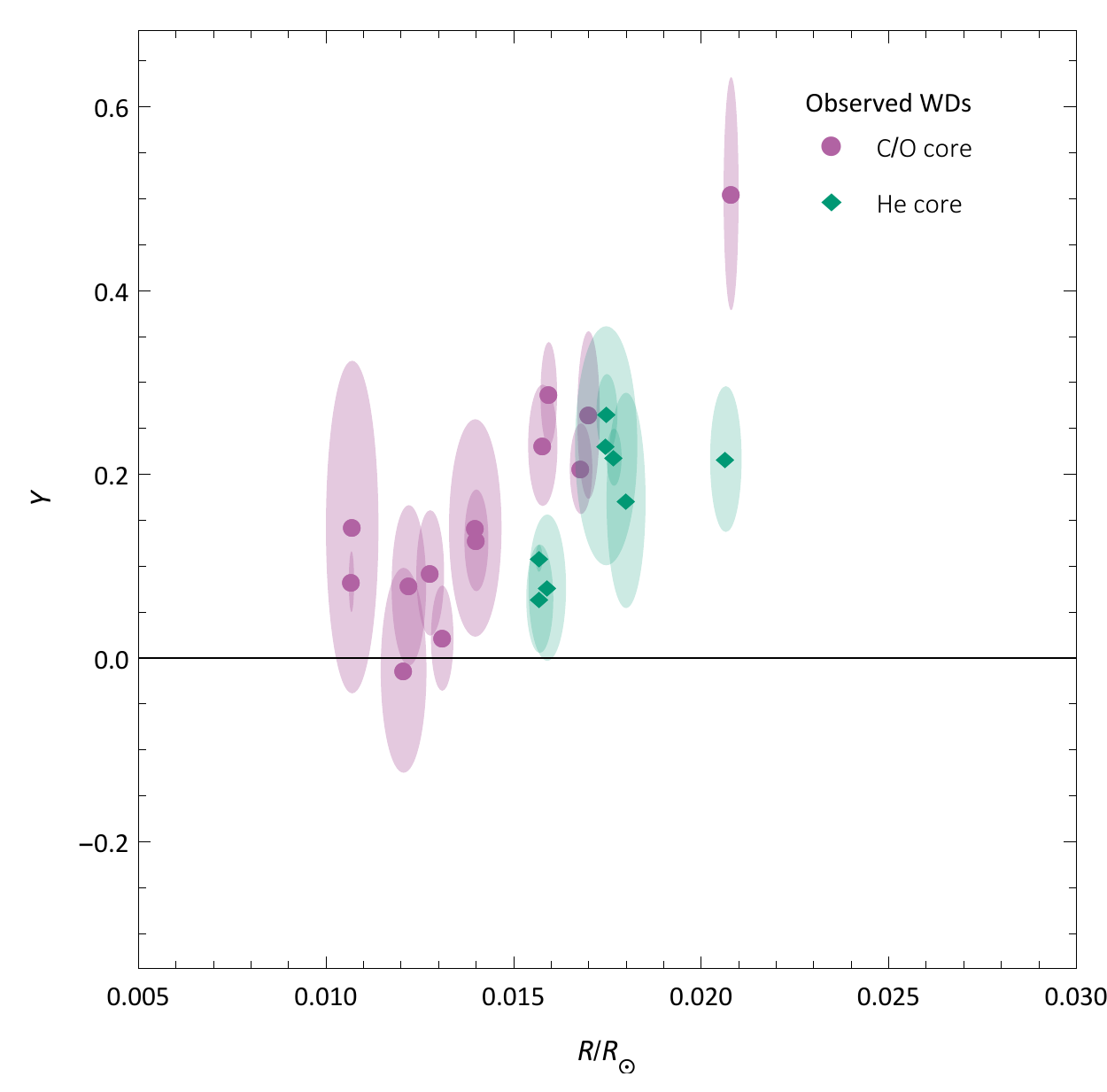}}
	\subfloat[Thick envelopes\label{plot:YRthick}]
	{\includegraphics[width=\columnwidth]{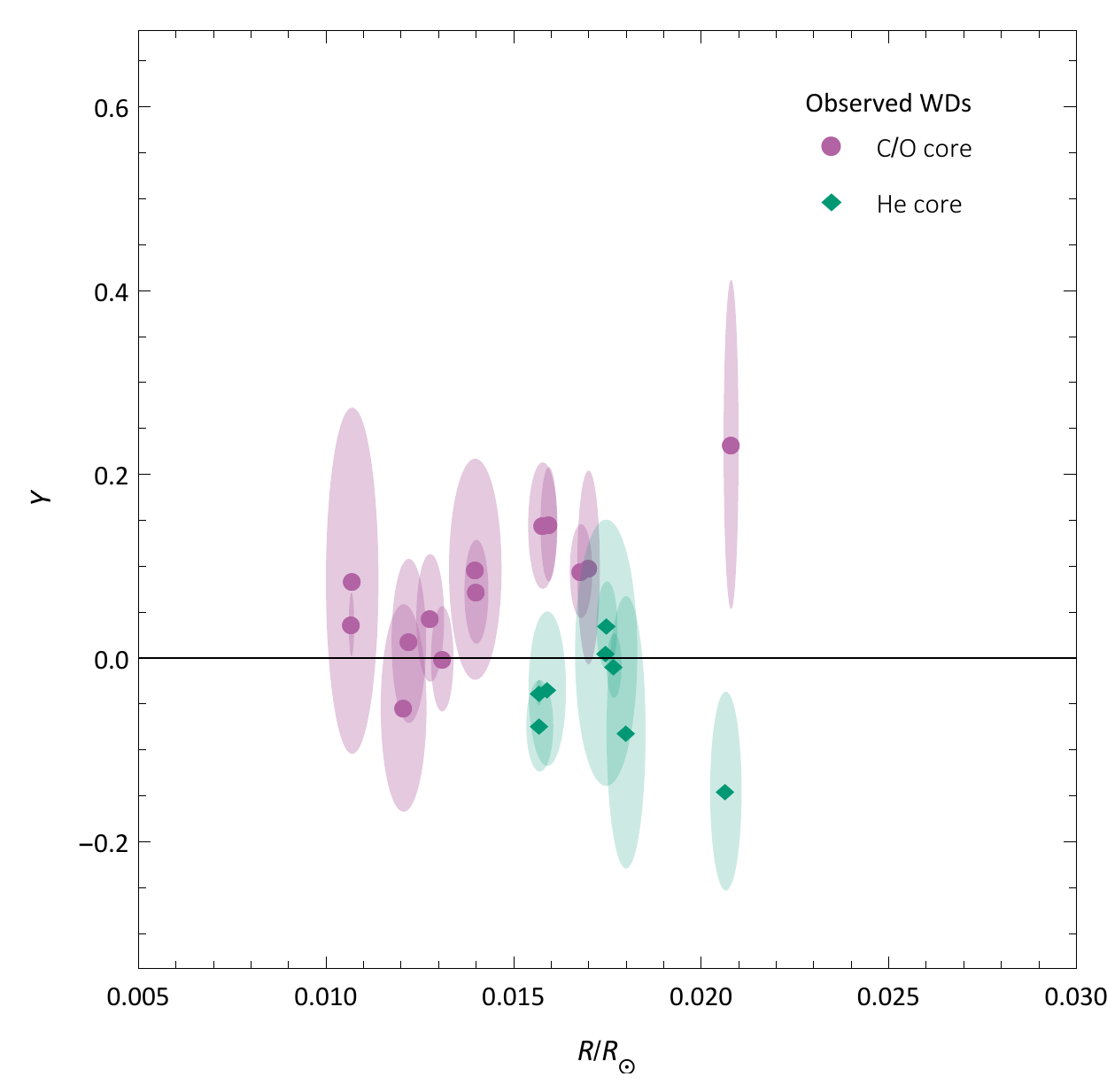}}
	\caption{Preferred central values of the modified-gravity parameter $Y$ along with their 1$\sigma$ uncertainty against observed WD radii. We assume a HS equation of state corrected for effective temperature and a thin/thick envelope (see Section~\ref{sec:Temp/Env}). Most notably, this removes the spurious correlation of Fig.~\ref{plot:YRT} and gives a dataset internally consistent for constraining gravity. Thin envelopes appear to only be consistent with modified gravity ($Y \neq 0$), while thick ones bring the stars into consistency with GR.}
\end{figure*} 
The magnitude of the effect of temperature on He vs C/O cores is very different, and simulations are not available in the same ranges of mass and temperature. For this analysis, we thus use the identification of the cores as per Ref.~\cite{Parsons2017}, identifying the stars as either He, or C/O, where for the latter we always marginalise over $Z=6,8$ with equal probability. Moreover, the available simulations for He have a maximum temperature of $T=2\cdot10^{4}$~K, so we exclude from the sample the 5 stars identified as having larger effective temperatures. For each star, we also fix the effective temperature in the model to its central observed value. The uncertainties in the observed temperatures are similar to those in the mass and radius, but the effect on the radii is a small additional correction which would not affect the end result, but would significantly increase the computational requirements. 

The properties of the envelope (thickness) should also be treated as a free unobservable parameter, local to each star, which would require marginalization, unless asteroseismological data were available to constrain it (see Section~\ref{sec:Temp/Env}). However, we only have access to two series of simulated mass-radius relations with envelopes for each core composition. We thus perform the analysis separately on each of the He- and C/O-identified sets of stars for each of the two types of envelopes: thick and thin. In the mass-radius series with the thin envelopes, for He stars, there is no envelope in the simulations we use. For C/O stars, the thin envelope does not contain hydrogen, but only helium; even though the helium fraction is fixed, changing it does not significantly affect the properties of the WDs \cite{Romero2012}. The thin envelopes therefore model the smallest effect an envelope could have. On the other hand, the mass-radius series with the thickest envelopes that are publicly available ($M_\text{H}/M=3\cdot 10^{-4}$ for He WDs and $M_\text{H}/M=10^{-5}$ for C/O cores from Ref.~\cite{Panei:1999ji}) comprise envelopes with hydrogen mass fractions one order of magnitude thinner that the maximal envelopes which can be produced in simulations for the relevant mass range ($4\cdot 10^{-3}$ for He-core WDs in Ref.~\cite{Benvenuto1998}, $2\cdot 10^{-4}$ for C/O-core WDs in Ref.~\cite{Romero2012}). This allows us to learn the significance of the effect of increasing the amount of hydrogen in the envelope within modifications of gravity.%
\footnote{Notice that, knowledge of its extreme value, could certainly be used to provide a hard lower bound for $Y$.}%

In Fig.~\ref{plot:YRthin}, we show the central $Y$ values preferred by each star on the assumption of a finite-temperature-corrected equation of state with a thin envelope, according to the prescription detailed in Section~\ref{sec:Temp/Env}. This is to be contrasted with Fig.~\ref{plot:YRT}. The addition of finite temperature/envelopes in the model removes the spurious correlation of $Y$ with radius/temperature and makes the dataset internally self-consistent for the purpose of constraining $Y$. Nonetheless, there is still a tendency for the stars to prefer a positive $Y$, a feature removed by further modelling. 

Fig.~\ref{plot:YRthick} shows the change in the preferred values of $Y$, but for a thick hydrogen envelope choice. Compared to the thin envelope, the overall preferred value of $Y$ decreases, as thicker envelopes yield a larger radii at constant mass and effective temperature. As we will discuss in Section~\ref{sec:Yconstr}, the data for the two types of cores under the assumption of the thick-envelope parameters is consistent with GR, albeit marginally inconsistent with each other.\\

\paragraph*{Polytrope:} 

Before we close this section let us make a comment on the {\it polytropic} approximation to the equation of state. The polytropic approximation is a very useful approximation allowing to study the gravitational dynamics of the star in a simple setup, parametrising the microphysics of the interior in terms of the dimensionless polytropic index $n$. For applications of it in a context similar to ours see for example \cite{DavisLim2012, SaksteinOscil,SaksteinJain2014,Jain:2015edg, Koyama}. Performing a similar statistical analysis with the one outlined earlier, we found that the constraint on the polytropic index is completely degenerate with that on the modified gravity parameter $Y$, i.e.\ the likelihood maximises along an infinite stripe on the $Y$-$n$ plane. More precisely, a change in $n$ at the level of 1\%, can be compensated for by a change in $Y$ of order 1. Although the range of the polytropic index $n$ is constrained on physical grounds, the strong sensitivity of $Y$ on $n$ means that any constraint on the former is essentially dictated by the (theoretical) prior assumed for $n$. This is especially problematic since the values of $n$ preferred by each star differ by multiple $\sigma$, implying that, for this data set, the polytropic index is a local feature, also related to the temperature of the star. This is an important weakness of the polytropic approximation when constraining gravity, and therefore we will not consider it any further here.

\section{A new bound on the gravity modification $Y$\label{sec:Yconstr}}
In the previous section we discussed the statistical analysis for three different interior structures when modelling the WDs in the dataset: the cold HS model, and the finite temperature ones with a thin and thick hydrogen envelopes. Each case provides some information about the possible gravity modification in the stars' interior. 

\begin{figure}[htb]
	\includegraphics[width=\columnwidth]{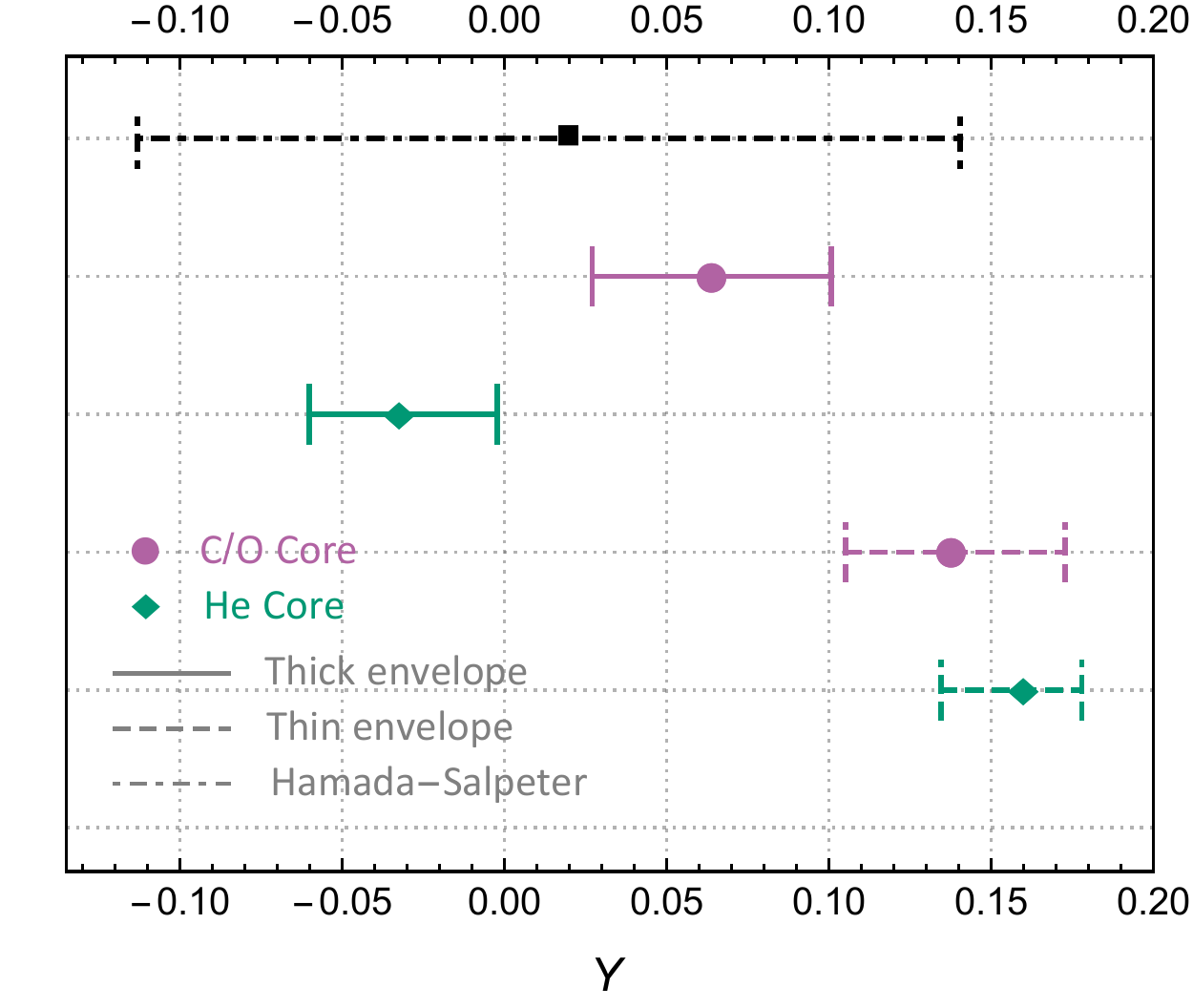}
	\caption{The marginalized combined constraints on the dimensionless modified gravity parameter $Y$ at $2\sigma$ from various subsets of the WD catalog made internally consistent by considering different models for the equation of state and envelope as described in Section \ref{sec:Yconstr}. Dashed and continuous lines correspond to finite temperature models with a thin and thick envelope respectively, while the dot-dashed bar corresponds to the constraint from the coldest star modelled according to a (zero temperature) HS equation of state in modified gravity. Remember that, the latter HS constraint is derived after marginalising over helium and C/O cores. The tightest and robust upper bound at $2\sigma$ comes from the coldest star modelled within HS, yielding $Y < 0.14$.}\label{plot:Ymarg} 
\end{figure}

The result from the cold HS case, though dominated by systematics, can still be used to provide an upper bound on $Y$: temperature and envelope always act as to increase the radius of the star for a given mass, and so does weakening gravity through a $Y > 0$. The two are at least to some extent degenerate and accounting for temperature effects can move stars only towards smaller $Y$ in Fig. \ref{plot:YRT}. Whatever the model for gravity, it must be compatible with all the observed stars, and in particular with the coldest star. Therefore, the coldest star in Fig. \ref{plot:YRT} (SDSS J0138-0016) defines the most conservative {\it upper bound} on $Y$ given the dataset, which is 
\begin{align}
Y &< 0.14 \,\,\text{at }2\sigma\, , \label{Yupper}\\ 
Y &< 0.19 \,\,\text{at }5\sigma\,. \notag 
\end{align}
Since we have fully taken into account the effects of modified gravity in the HS equation of state, this is a hard upper bound for $Y$.\\

We now turn to the constraints on $Y$ derived in the presence of finite temperature/envelope corrections according to the procedure outlined in Section \ref{sec:data}, for the distinct cases characterised by the different envelope/core structure assumptions.

\begin{enumerate}

\item[a.] \emph{Thin envelopes} (Helium-core stars with no envelope and C/O-core ones with a helium envelope): From the combined likelihood for the 21 stars we have retained in the sample, we find that at 2-$\sigma$ confidence level, for Helium-core stars it is  $0.13 < Y < 0.18$, %
and similarly for C/O-core ones we find that $0.11 < Y < 0.17$. %
Notice that, the preferred value of $Y$ is now consistent amongst the whole set of stars. This constraint is telling us that for the given thin envelope structure, both types of stars prefer a modification of gravity with a strength weaker than GR.
 
Most importantly, the He-core stars in this analysis are assumed to have no envelope whatsoever. This yields the smallest possible configurations they can have at finite temperature, in turn providing a handle upon an upper bound for $Y$. We can thus quote a new upper bound from multiple He stars
\begin{align}
	Y &< 0.178 \,\,\text{at }2\sigma\, , \label{YupperHe}\\
	Y &< 0.182 \,\,\text{at }5\sigma\,. \notag 
\end{align}
In any case, it can be said with certainty that the dataset is incompatible with {\it both} thin envelopes and GR simultaneously, both for He and C/O stars.

\item[b.] \emph{Thick envelopes} (Helium-core stars with hydrogen, and C/O-core ones with a helium-hydrogen envelope respectively): Under a similar logic as before, the corresponding 2-$\sigma$ intervals turn out to be $-0.06 < Y  < -0.002$ for He-core stars and $0.027 <Y < 0.10$ for C/O ones respectively.
The first point to be made is that, the preferred values of $Y$ have been shifted to the left compared to the thin-envelope modelling, as expected from the increase in the envelope's thickness, with He-core stars preferring a stronger gravity than GR ($Y < 0$). Though the results are not strictly compatible with GR, they could be made so with a small change in the envelope's thickness, they certainly reflect the high improvement in our modelling which is to be regarded significantly more robust than any constraints that would be placed with the simple cold, Fermi/polytropic model.%
\footnote{Notice that, the thick-envelope analysis would be able to provide a hard lower bound to the value of $Y$ if we were certain that the envelope thickness is the \emph{maximum} that is achievable in simulations. Unfortunately, as explained in Section~\ref{sec:Temp/Env}, it has been shown that both the C/O and He simulations can result in envelopes with mass fractions in hydrogen one order of magnitude higher than the ones we had access to. These envelopes would inevitably result in larger radii and therefore move the bounds on $Y$ to lower values.}

As a corollary of the above analysis, one can safely make the following statement: The helium-core stars in the WD catalog of Ref.~\cite{Parsons2017} cannot have a hydrogen envelope with mass fraction higher than $3\cdot 10^{-4}$ and still remain compatible with GR predictions. At the same time, the C/O stars in the catalog must have an envelope with mass fraction in hydrogen of at least $10^{-5}$ if they are to be compatible with GR.
\end{enumerate}
We have presented the individual constraints on the combination plot~\ref{plot:Ymarg}. Combining the posteriors, we can conclude that the data places an {\it upper bound} of
\begin{align}
	Y &< 0.14 \,\,\text{at }2\sigma\, , \label{YupperCombined}\\
	Y &< 0.18 \,\,\text{at }5\sigma\,. \notag 
\end{align}

This bound is, to our knowledge, the strongest in the literature for the new physics encoded in the modified-Poisson equation \eqref{Poisson}. For the case of the \emph{Beyond Horndeski} model, corresponding to a particular subclass of the models captured by our analysis, it has been shown that after the determination of GW speed by LIGO, the parameter $Y$ also determines the value of the Newton's constant in screened regions \emph{outside} compact objects. This effect was used to constrain $Y$ for this particular model, without reference to internal structure of the stars, but with the use of the decay of the orbit of binary pulsars. The excellent bound thus obtained, $Y<7.5\cdot 10^{-3}$ \cite{Dima:2017pwp}, is nonetheless weakened once more general models are considered, since the two physical effects are no longer controlled by the same parameters. Our bound thus still serves to improve the constraints in DHOST models of gravity and provides a completely orthogonal constraint.

Ref.~\cite{Jain:2015edg} used the WD mass-radius observations of \cite{holberg} and an idealised cold Fermi equation of state, to find the 2$\sigma$ (5$\sigma$) constraint of $Y<0.27$ ($Y< 0.54$). We are thus presenting a three-fold improvement in the upper bound, driven to a large extent by significantly reduced uncertainties in the data and our improved modelling of systematic effects. Ref.~\cite{Jain:2015edg} also placed a lower bound $Y>-0.22$ at 2$\sigma$ by requiring that the Chandrasekhar mass be larger than the heaviest WD observed. However, it was shown in Ref.~\cite{Babichev:2016jom} that under the inclusion of relativistic corrections the value of the Chandrasekhar mass changes significantly for nagative $Y$, weakening this bound to $Y>-0.48$.%
\footnote{Note that, in the relativistic regime spherical solutions exist only for $Y>-0.44$ (to be contrasted with the non-relativistic $Y > -0.67$), as explained in Section~\ref{sec:Hydro}.} 

Another constraint was found by demanding that the lightest observed red dwarf is at least as heavy as the minimum mass for the onset of hydrogen burning \cite{Sakstein:2015zoa,Sakstein:2015aac}. This leads to a constraint that $Y<1.6$, although this is dependent on having assumed a polytropic equation of state for the red dwarf with a fixed index, $n$. Our constraint is thus an order of magnitude better than that obtained from red dwarfs. We should stress here again though, that our analysis for the polytropic equation of state in WDs, revealed a degeneracy between $Y$ and $n$, implying that this equation of state could not be used to constrain such modifications of gravity.

Finally, in Ref.~\cite{Sakstein:2016ggl}, modifications of gravity are constrained from the analysis of weak lensing and X-ray profiles of 58 galaxy clusters with an averaged redshift of 0.33, obtaining $Y=-0.11^{+.93}_{-0.67}$.

The fact that two independent methods give us relatively close values of $Y$ is a guarantor of the quality of these two upper $Y$ bounds.
Nevertheless, since the models of white dwarfs (with and without modified GR)  used in  this analysis describe accurately the structure 
of the different white dwarfs, as well as the observational data,  this gives us some certainty about the robustness of the upper limit found in this study.

%
%
\section{Discussion and summary}
The study of stars as laboratories for testing the standard gravity paradigm provides an exciting and powerful way to understand and constrain alternative gravitational theories. 
Here, we considered the whole family of scalar-tensor models beyond GR, still viable after GW170817, which make a sharp prediction: a fifth-force effect occurs in the star's interior, but not outside, so constraining the mass-radius relation provides an excellent probe, while typical tests performed in the solar system would not be informative. In compact massive bodies, this type of fifth force can act to weaken gravity --- in fact, one of the implications of the recent constraints on the speed of gravitational waves is that any modification of gravity that reduces its strength must have exactly the behaviour inside objects that we are constraining in this paper \cite{Amendola:2017orw}.

For our analysis, we employed the most recent and accurate compilation of independent mass and radius observations of WDs in binaries. In constraining alternative gravity theories at stellar scales the usual choice has been the Fermi equation of state or its polytropic approximation. Until now, it was usually to some extent justifiable to neglect temperature effects in this context. However, the uncertainties in the catalog we used \cite{Parsons2017}, are significantly smaller than previously available (e.g.\ \cite{holberg}) and we have shown that ignoring either finite-temperature effects and envelope structure leads to a very significant systematic error and potentially false conclusions about the gravitational law. What is more, we found that within the polytropic approximation of the (cold) Chandrasekhar model, also widely used in the literature, the effect of modified-gravity parameter is in complete degeneracy with the polytropic index, making it a rather unsuitable choice for the purpose of testing gravity, at least using WDs. 

Driven by the above, we modelled finite temperature and envelope corrections on top of the cold Hamada-Salpeter model for WDs using publicly available data from stellar evolution simulations to estimate the size of the possible corrections, and studied the effect of the fine structure of the star's interior in the presence of a modification of gravity. To our best of knowledge, this is the first time this is pursued in the context of theories beyond GR. Our analysis revealed a degeneracy between the stars' structural properties and modified gravity: in principle, observations of masses/radii can always be reproduced through an appropriate balance between the star's temperature, envelope thickness and a modification of gravity. This comes to add to the already known degeneracies, such as the one between the star's pulsation frequency and envelope structure. 
Nonetheless, we were still able to place a hard upper bound on the gravity modification $Y$,
\begin{align}
Y \leq 0.14\,,
\end{align}
at 2$\sigma$, an improvement by a factor of three compared to previous bounds in the literature. At the same time, our modelling of the star's interior eliminated a significant amount of systematics, strengthening the robustness of the above constraint. 
We cannot place a lower limit on $Y$ since the thickest envelopes in the public datasets are one order of magnitude thinner than is possible to create in simulations. Access to mass-radius simulation results with thicker envelopes would allow us to place a stringent lower bound on $Y$.

Proper modelling of the stars' evolution requires the simulation of the whole evolutionary track from its progenitor, in the context of modified gravity: energy transport must be modelled, which does not allow us to neglect the time dependence of the stellar solution. Modifying the existing public codes to understand the interplay between finite temperature and modified gravity, is the natural next step from this work. Even without modelling the envelopes in He-core WDs, it should be possible to remove any remaining systematic contribution to our upper bound from the effect of the interaction of temperature and modified gravity.

It should be apparent from this work that the lack of measurement of the thickness of the envelope plays the ultimate limit in our ability to constrain gravity using white dwarfs. The existing measurements of the variability timescales of ZZ Ceti stars from astreroseismological observations show that it is possible to constrain the envelope thickness. A pulsating white dwarf with an independently measured mass and radius might therefore provide the ultimate tool to constrain modifications of gravity. We leave understanding the details of these physics for future work.

\begin{acknowledgments}
I.S.~and I.D.S.~are supported by European Structural and Investment Funds and the Czech Ministry of Education, Youth and Sports (Project CoGraDS --- CZ.02.1.01/0.0/0.0/15\_003/0000437). I.L.~thanks the Funda\c c\~ao para a Ci\^encia e Tecnologia (FCT), Portugal, for the financial support to the Multidisciplinary Center for Astrophysics (CENTRA),  Instituto Superior T\'ecnico,  Universidade de Lisboa (Grant No. UID/FIS/00099/2013).
\end{acknowledgments}

\bibliographystyle{utcaps}
\bibliography{WD-Bib}

\end{document}